\documentclass[runningheads]{llncs}

\usepackage{url}
\usepackage{enumitem}
\usepackage{multirow, graphicx}
\usepackage[dvipsnames]{xcolor}
\usepackage{booktabs}
\usepackage{array}
\usepackage{lmodern}
\usepackage[T1]{fontenc}
\usepackage[normalem]{ulem}
\newcommand\orangesout{\bgroup\markoverwith{\textcolor{orange}{\rule[0.5ex]{2pt}{0.4pt}}}\ULon}
\usepackage{framed}
\usepackage{colortbl}
\usepackage[utf8]{inputenc}
\usepackage{color, colortbl}
\definecolor{Gray}{gray}{0.9}
\usepackage{caption}
\usepackage{multirow,array}
\usepackage[utf8]{inputenc}
\usepackage[misc,geometry]{ifsym}

\begin{document}
\sloppy
\title{Developers' perception on the severity of test smells: an empirical study}

%
%

\author{Denivan Campos\inst{1} \and Larissa Rocha\inst{1,2} \and Ivan Machado\inst{1}}

\authorrunning{Campos et al.}
%
\institute{Institute of Computing, Federal University of Bahia (UFBA), Salvador - BA, Brazil \and State University of Feira de Santana (UEFS), Feira de Santana - BA, Brazil \\ \email{denivan.campos@ufba.br, larissa@ecomp.uefs.br, ivan.machado@ufba.br} }

\maketitle              

\begin{abstract}
Unit testing is an essential component of the software development life-cycle. A developer could easily and quickly catch and fix software faults introduced in the source code by creating and running unit tests. Despite their importance, unit tests are subject to bad design or implementation decisions, the so-called test smells. These might decrease software systems' quality from various aspects, making it harder to understand, more complex to maintain, and more prone to errors and bugs. Many studies discuss the likely effects of test smells on test code. However, there is a lack of studies that capture developers' perceptions of such issues. This study empirically analyzes how developers perceive the severity of test smells in the test code they develop. Severity refers to the degree to how a test smell may negatively impact the test code. We selected six open-source software projects from GitHub and interviewed their developers to understand whether and how the test smells affected the test code. Although most of the interviewed developers considered the test smells as having a low severity to their code, they indicated that test smells might negatively impact the project, particularly in test code maintainability and evolution. Also, detecting and removing test smells from the test code may be positive for the project.

\end{abstract}

\section{Introduction}

Software testing is an essential strategy to ensure the quality of software systems. Creating and executing test code encompass activities that require a lot of effort and cost, especially when manually developed \cite{peruma2019distribution}. Nonetheless, the need to find defects early in the development cycle has increased the interest in employing unit testing practices \cite{breugelmans2008testq}.

In recent years, test code has become persistent and needs to evolve along with the production code to be more effective \cite{breugelmans2008testq}. For example, at Google, the amount of code and the size of the test pool grows linearly to maintain the software quality. Google runs over 2 Million Lines of Code (LOC) and 150 million tests per day \cite{memon2017taming}. Keeping up this whole process of evolving and maintaining these test cases might be a time-consuming task. In addition, as the amount of tests grows, the effort invested in maintaining the tests can make it a costly activity. On the other side, creating and running unit tests can quickly capture and support fixing software failures introduced in the source code \cite{peruma2019distribution}.


Despite their importance, unit tests might be subject to inappropriate design or poor implementation decisions. Such issues may lead to the insertion of smells in test code, the so-called test smells \cite{bavota2012empirical}. The presence of test smells makes test code more difficult to comprehend, more complex to maintain, and prone to errors \cite{peruma2019distribution,spadini2018relation}. Software developers and testers are usually unaware of the concept of test smells \cite{silva2020survey}.


Software projects commonly develop extensive sets of test cases to assure the quality of developed systems \cite{reichhart2007rule}. To this effect, automated support for creating test cases has gained momentum, and the software testing community has released many support tools lately \cite{fraser2011evosuite,pacheco2007randoop}. In addition, many tools are providing automated strategies to prevent, detect, and refactor test smells, as discussed by Aljedaani et al. \cite{tsdetectiontoolsAljedaani}.



Aligned with the need to understand whether and how harmful test smells are to a software project, many researchers have empirically evaluated the likely effects of test smells \cite{garousi2018smells,meszaros2007xunit,silva2020survey,spadini2020investigating}. In a recent study, Garousi et al. \cite{garousi2018smells} enlisted the main negative consequences of the presence of test smell in test code: smells reduce tests changeability, stability, readability, and maintainability. Silva-Junior et al. \cite{silva2020survey} surveyed software testing professionals to understand and analyze how often they insert test smells into the test code and the reasons behind it. The authors showed that, independently of the professional experience, they are prone to insert test smells in their test code, even when using standard practices.


Although some studies claim that test smells might hinder code maintenance, there is still a lack of studies that capture the developers' perception of that issues. In this context, this study investigates how developers recognize test smells. We aim to empirically analyze how developers perceive the severity of test smells in the test code they develop. We refer \textit{severity} to the degree to which a given test smell can negatively impact the test code. In addition to severity, we were also interested in investigating how the test code would behave after removing the test smells.

We conducted an empirical study split into three phases. In the first, we selected six open-source software projects from GitHub. The second phase consisted of interviews with the developers of the selected projects to investigate their perception of test smells. During the interviews, we introduced the definitions of each test smell and asked the developers whether they could refactor them. The third step was the analysis of the gathered data. To support this empirical study, we used the JNose Test \cite{Tassio2020Tools} tool to identify the test smells in the selected projects.

In summary, our study brings the following contributions:
\begin{itemize}
    \item an empirical study showing that most developers perceive test smells as low or middle in severity for their software projects;
    \item initial insights of the consequences of the test smells after refactoring. 
\end{itemize}

\section{Test Smells}

Test smells are sub-optimal designs in the unit test code that the developer chose when implementing test cases \cite{tsdetectiontoolsAljedaani,garousi2018smells,palomba2016diffusion,peruma2018smell}. Deursen et al. \cite{van2001refactoring} presents a catalog of 11 test smells and proposed refactoring strategies to remove them from the test code. Peruma et al. \cite{peruma2018smell} extended that catalog and included another 12 new test smells inspired by bad practices in unit test-based programming. Other researchers identified other test smells types and analyzed the effects of various types on production and test code  \cite{meszaros2007xunit,meszaros2003test,peruma2019distribution,spadini2018relation,virginio2019coverage}.

In this study, we analyzed eight types of test smells. We selected the most frequent ones in the test suites of the projects used in the study, which we briefly introduce next:

\begin{itemize}
    \item \textbf{Assertion Roulette (AR):} 
    It occurs when a test method contains more than one assertion statement without an explanation or parameter in the assertion method \cite{van2001refactoring};
    
    \item \textbf{Eager Test (ET):}
   When a test method contains multiple calls to multiple production methods \cite{van2001refactoring};

    \item \textbf{Empty Test (EpT):} 
    It occurs when a test method does not contain executable instructions \cite{peruma2018smell};

    \item \textbf{Lazy test (LT):} 
    It occurs when multiple test methods call the same production method \cite{van2001refactoring};

    \item \textbf{Redundant Assertion (RA):} 
    When the test methods contain assertion statements that are always true or always false \cite{peruma2018smell};

    \item \textbf{Constructor Initialization (CI):} 
    It occurs when a test method implements a constructor. It is recommended that all the fields be initialized inside the \textit{setUp()} method \cite{peruma2018smell};
    
    \item \textbf{Unknown Test (UT):}
    When a test method does not contain assertions. An assertion statement describes an expected condition for a test method \cite{peruma2018smell};\looseness=-1

    \item \textbf{Sensitive Equality (SE):} 
    It occurs when the test method makes an equality checking using the \textit{toString} method \cite{van2001refactoring}.  
  
\end{itemize}

\subsection{JNose Test Tool}

Aljedaani et al. \cite{tsdetectiontoolsAljedaani} identified 22 tools for test smell detection. Most of the tools support the JUnit framework and Java programming language. For our study, we applied the JNose Test \cite{Tassio2020Tools}, as it elaborates on the state-of-the-art tool by providing users with an easy-to-use graphical interface that facilitates the detection of test smells. It analyzes Java projects that use Maven and JUnit Framework\footnote{\url{https://junit.org/}}.\looseness=-1
 
The JNose Test\footnote{\url{https://github.com/arieslab/jnose}} is a web-based application that analyzes the quality of the test code by detecting test smells in the Java test code \cite{Tassio2020Tools}. JNose Test detects 21 types of test smells by following the detection rules from the TsDetect tool \cite{peruma2020tsdetect}. The tool encompasses a set of rules to identify and quantify the types of test smells in each test class and supports the analysis of test smells through several project versions \cite{Tassio2020Tools,virginio2020empirical}. Once started, the tool requires the user to configure the data entry to enable and specify one of the four types of analysis mode, \textit{TestClass}, \textit{TestSmell}, \textit{TestFile}, and \textit{Evolution}. 
JNose shows, as a result, the amount of LOC, methods, types, and amount of test smells in each test class of the project. With such a tool, it is possible to verify the test quality from an evolutionary perspective. It enables capturing metrics, and the occurrence of test smells through several project versions.

\section{Research Methodology}
\label{methodology} 


We defined two research questions in this study:

\begin{enumerate}[label=\bf RQ\arabic*. ,leftmargin=1.1cm]

  
     \item  \textbf{How do software developers consider the severity of test smells and their effects on test code quality?} We aim to analyze how the software developers perceive the impact of the presence of test smells in test code, in terms of quality attributes such as comprehensibility and maintainability.


    \item \textbf{What is the behavior of test code after refactoring test smells?} We aim to analyze the changes in the test code after refactoring test smells.\looseness=-1

\end{enumerate}

\subsection{Study Design}

\begin{figure}[ht]
\centerline{\includegraphics[width=\linewidth,height=\textheight,keepaspectratio]{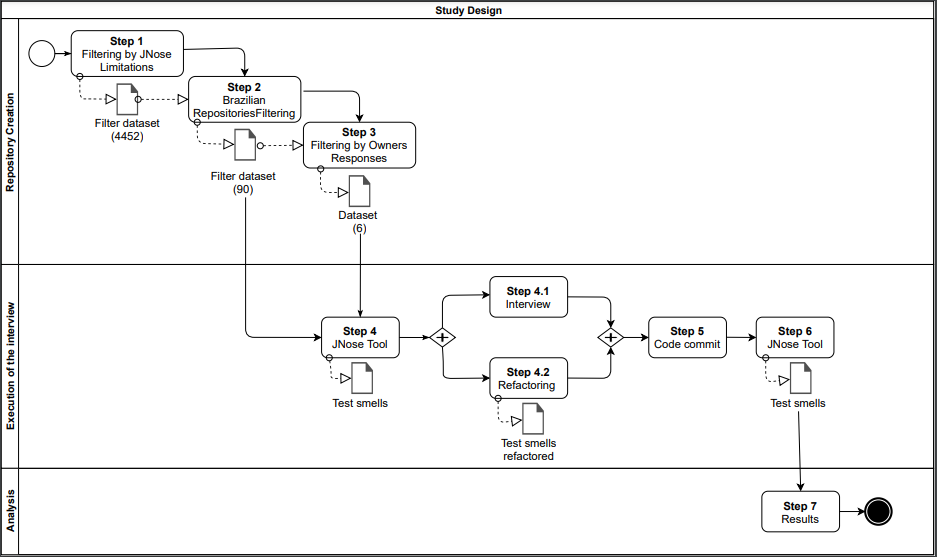}}
\caption{Empirical Study Design} 
\label{stepsstudy}
\end{figure}

The design employed in this study consists of three phases: \textit{Repository Creation}, \textit{Interview}, and \textit{Data Analysis}. Each phase comprises a set of steps, as Figure \ref{stepsstudy} shows.\looseness=-1 

\vspace{0.5cm}
\noindent\textbf{Phase 1 - Repository Creation}. This phase comprised the following steps:

\begin{itemize}
    \item \textbf{Step 1}. \textit{Filtering by JNose Test limitations.}
    In this step, we used a dataset composed of 21,482 projects available in public GitHub repositories. First, we manually filtered each of the repositories with the following criteria: projects written in Java, composed of at least two stars, with Issues, and test cases created with the JUnit framework (versions 4 or 5). The choice for projects written in Java and tests with JUnit was due to the JNose Test limitations. As a result, we obtained a sample with 4,452 public repositories.

    \vspace{0.2cm}
    \item \textbf{Step 2}. \textit{Filtering repositories by ownership.}
    In this step, we collected the URLs of the 4,452 projects and manually opened each one. We filtered the projects that contained e-mails with Brazilian developers. Then, we ran each of these projects in the JNose Test to validate them. As a result, we obtained a sample with 90 public repositories. 
    
    \vspace{0.2cm}
    \item \textbf{Step 3}. \textit{Filtering repositories by responses}.
    In this phase, we sent out e-mails to the 90 selected projects. We received six positive replies, which composed our dataset.\looseness=-1
    
\end{itemize}

\noindent
\textbf{Phase 2 - Interview}.
The interview considered each of the developers individually. It comprised the following steps:

\begin{itemize}
    \item  \textbf{Step 4}. \textit{Executing the JNose Test}. The first step was to execute the selected projects in the JNose Test  \cite{Tassio2020Tools} to gather evidence about the test smells identified in each project. Based on the yielded results (e.g., number of test smells present in each project), we created the protocol for the interviews. 

    \vspace{0.2cm}
    \item \textbf{Step 4.1}. \textit{Carrying our the interviews}.  
    Next, we proceeded with the actual interviews. First, we carried out the characterization of the participants, which included gathering information on their knowledge about creating and maintaining test cases. We created an online questionnaire using Google Forms and sent it to the interviewees. The interviews were held virtually, through Google Meet, and guided by the protocol created after detecting test smells in each of the projects. We presented the JNose Test to the participants during the interviews, explained that we ran the tool in their project, and presented the test smells we detected in each project. We also showed the most frequent test smells. 

    \vspace{0.2cm}
     \item \textbf{Step 4.2}. \textit{Refactoring test code}. 
     After analyzing the presence of test smells in the test code, they had to discuss (i) whether the test smells were harmful to their projects and (ii) whether they should refactor them. In this step, the developers had to refactor at least one test smell. We did not impose which test smell they must refactor. They could select any test smell they wanted.
     
      \vspace{0.2cm}
     \item \textbf{Step 5}. \textit{Commiting test code changes}.
     After refactoring, we asked the developers to commit the changes to their GitHub repositories. This step was essential to allow us to rerun the JNose Test on the refactored project. 
      \vspace{0.2cm}
      
     \item \textbf{Step 6}. \textit{Rerunning the JNose Test}.
    In this final step, we reran the project in the JNose Test to compare the test classes before and after refactorings.
\end{itemize}

\noindent
\textbf{Phase 3 - Data Analysis.} This phase comprises the analysis of gathered data, as follows:
\begin{itemize}
    \item \textbf{Step 7}. \textit{Results}. In this step, we present the results of our empirical study. This research has an exploratory character with a qualitative analysis. For open-ended questions that could combine multiple answers, the sum of percentages could be greater than 100\%. 
\end{itemize}

\section{Results} \label{results}

\subsection{Developers' Profiles}

In this study, we interviewed the developers of each selected project. We next describe their background.

They all hold a Bachelor's degree (two in Computer Engineering, two in Electrical Engineering, and two in Business Information Systems), and two out of them have a Master's degree. They work in different Brazilian states: three of them work in São Paulo, and the others work in the states of Rio Grande do Sul, Bahia, and Santa Catarina.

Regarding their professional experience with software development, three had over ten years of professional experience, two had less than ten years, and one developer had less than five years. Among the three developers with more than ten years of experience, two of them also had more than ten years of experience in software testing. Among the two developers with up to 10 years of development experience, one also had a strong background in software testing. The others have between one and five years of experience in software testing.

\subsection{Characterization of the selected projects}

The six projects we analyzed vary in size - considering the number of LOC and the number of test methods, as Table \ref{tab:projectos} shows. The largest project is the \texttt{Fatiador} (the project aims to convert flat strings to Java objects and vice versa) with 162 test methods.
For the \texttt{csv2Bib} (the project aims to convert \texttt{.csv} files into \texttt{.bib} or \texttt{.ris} files) and \texttt{Our Digital Bank} (the project provides an API to support the operation of a our digital bank) projects, we could only select two test smell types each.
 
\begin{table}[t]
    \scriptsize
\def \arraystretch{1.2}
  \centering
  \caption{Selected projects}
  \begin{tabular}{p{11em}cccl}
    \toprule
    \multicolumn{1}{c}{\textbf{Project}} & \multicolumn{1}{c}{\textbf{LOC*}} & \multicolumn{1}{c}{\hspace{.5em}\textbf{Test Methods}\hspace{.5em}} & \multicolumn{1}{c}{\hspace{.5em}\textbf{Test Classes}\hspace{.5em}} & \multicolumn{1}{c}{\hspace{.5em}\textbf{Project URL}\hspace{.5em}}  \\
    \midrule
    \textit{Our Digital Bank}      & \centering 84    & 7   &   4 & \url{https://bit.ly/3ABZui2}\\
    \textit{csv2bib}               & \centering 98    & 5   &   2 & \url{https://bit.ly/3xlHyXf} \\
    \textit{CursoMassa}            & \centering 261   & 32  &   4 & \url{https://bit.ly/3dQEwSV}\\
    \textit{Dependency injection}  & \centering 390   & 23  &   5 & \url{https://bit.ly/36h3ff3}\\
    \textit{l2jserver}             & \centering 538   & 31  &   5 & \url{https://bit.ly/3hInK9I}\\
    \textit{Fatiador}              & \centering 2,362  & 162 &   17 & \url{https://bit.ly/36gtS3F}\\
    \bottomrule
    \multicolumn{5}{l}{(*) Number of LOC in the test classes.} 
    \end{tabular}%
  \label{tab:projectos}%
\end{table}%


\subsection{How do software developers consider the severity of test smells and their effects on test code quality? (RQ1)}

We made an attempt to analyze how the developers would perceive the impact of the presence of test smells in test code. To accomplish that, we first considered the degree of severity of each test smell. Table \ref{tab:severity} summarizes the data about the severity of the test smells identified in each project.

The \textit{AR} test smell was the one which varied the most in terms of severity. Three projects presented variations for this test smell. The project \texttt{Curso Massa} (this project aims to support programming classes) presented 2 degrees of severity for the \textit{AR} test smell: \textit{low} and \textit{middle}; the project \texttt{l2jserver} (this is a rewrite of the l2jserver project\footnote{https://bitbucket.org/l2jserver/}) also presented 2 degrees of severity: \textit{low} and \textit{high}; and the project \texttt{Dependency Injection} (this project implements a dependency injection framework in Java with annotations) presented 3 degrees of severity: \textit{low}, \textit{middle}, and \textit{high} severity. 

In addition, the \textit{UT} and \textit{ET} test smells also presented differences in their perceived severity, ranging from \textit{low} to \textit{middle} and \textit{low} to \textit{high}, respectively.
The remaining projects and test smells only presented a \textit{low} severity degree. 

Although most of the interviewees classified \textit{AR} test smell as \textit{low} severity, we observed small variations. For example, the developer of the project \texttt{Dependency Injection} analyzed 3 test smells of the \textit{AR} type and gave a different classification for each. For this developer, \textit{AR} could impact different her project in different ways. 


\begin{table}[t]
\scriptsize
  \centering
   \def\arraystretch{1.2}
   \caption{Degree of severity of test smells (TS) per project}
    \begin{tabular}{cc cc cc cc cc cc}
    \toprule
    
    \multicolumn{2}{c}{\textbf{Curso Massa}} & \multicolumn{2}{c}{\textbf{l2jserver}} & \multicolumn{2}{c}{\textbf{Dep. Injection}} & \multicolumn{2}{c}{\textbf{Fatiador}} & \multicolumn{2}{c}{\textbf{csv2bib}} & \multicolumn{2}{c}{\textbf{Our Digital Bank}} \\
     
     \midrule
     
    \rowcolor[rgb]{ .851,  .851,  .851} \multicolumn{1}{c}{TS} & \multicolumn{1}{p{1.3cm}}{\cellcolor[rgb]{ 1,  1,  1}\centering Severity} & 
    
    \multicolumn{1}{c}{TS} & \multicolumn{1}{p{1.3cm}}{\cellcolor[rgb]{ 1,  1,  1}\centering Severity} & 
    
    \multicolumn{1}{c}{TS} & \multicolumn{1}{p{1.3cm}}{\cellcolor[rgb]{ 1,  1,  1}\centering Severity} & 
    
    \multicolumn{1}{c}{TS} & \multicolumn{1}{p{1.3cm}}{\cellcolor[rgb]{ 1,  1,  1}\centering Severity} & 
    
    \multicolumn{1}{c}{TS} & \multicolumn{1}{p{1.3cm}}{\cellcolor[rgb]{ 1,  1,  1}\centering Severity} & 
    
    \multicolumn{1}{c}{TS} & \multicolumn{1}{p{1.3cm}}{\cellcolor[rgb]{ 1,  1,  1}\centering Severity} \\
    
    \midrule
    \rowcolor[rgb]{ .851,  .851,  .851} AR    & \cellcolor[rgb]{ 1,  1,  1}Low & AR    & \cellcolor[rgb]{ 1,  1,  1}Low & AR    & \cellcolor[rgb]{ 1,  1,  1}Middle & SE    & \cellcolor[rgb]{ 1,  1,  1}Low & AR    & \cellcolor[rgb]{ 1,  1,  1}Low & AR    & \cellcolor[rgb]{ 1,  1,  1}Low \\
    \rowcolor[rgb]{ .851,  .851,  .851} AR    & \cellcolor[rgb]{ 1,  1,  1}Low & AR    & \cellcolor[rgb]{ 1,  1,  1}Low & AR    & \cellcolor[rgb]{ 1,  1,  1}High & SE    & \cellcolor[rgb]{ 1,  1,  1}Low & AR    & \cellcolor[rgb]{ 1,  1,  1}Low & AR    & \cellcolor[rgb]{ 1,  1,  1}Low \\
    \rowcolor[rgb]{ .851,  .851,  .851} AR    & \cellcolor[rgb]{ 1,  1,  1}Middle & AR    & \cellcolor[rgb]{ 1,  1,  1}High & AR    & \cellcolor[rgb]{ 1,  1,  1}Low & SE    & \cellcolor[rgb]{ 1,  1,  1}Low & RA    & \cellcolor[rgb]{ 1,  1,  1}Low & EpT   & \cellcolor[rgb]{ 1,  1,  1}Low \\
    \rowcolor[rgb]{ .851,  .851,  .851} ET    & \cellcolor[rgb]{ 1,  1,  1}Low & ET    & \cellcolor[rgb]{ 1,  1,  1}Low & ET    & \cellcolor[rgb]{ 1,  1,  1}Low & AR    & \cellcolor[rgb]{ 1,  1,  1}Low & RA    & \cellcolor[rgb]{ 1,  1,  1}Low & EpT   & \cellcolor[rgb]{ 1,  1,  1}Low \\
    \rowcolor[rgb]{ .851,  .851,  .851} ET    & \cellcolor[rgb]{ 1,  1,  1}Low & ET    & \cellcolor[rgb]{ 1,  1,  1}High & ET    & \cellcolor[rgb]{ 1,  1,  1}Low & AR    & \cellcolor[rgb]{ 1,  1,  1}Low & CI    & \cellcolor[rgb]{ 1,  1,  1}Low & UT    & \cellcolor[rgb]{ 1,  1,  1}Middle \\
    \rowcolor[rgb]{ .851,  .851,  .851} ET    & \cellcolor[rgb]{ 1,  1,  1}Low & ET    & \cellcolor[rgb]{ 1,  1,  1}Low & ET    & \cellcolor[rgb]{ 1,  1,  1}Low & AR    & \cellcolor[rgb]{ 1,  1,  1}Low & CI    & \cellcolor[rgb]{ 1,  1,  1}Low & UT    & \cellcolor[rgb]{ 1,  1,  1}Low \\
    \rowcolor[rgb]{ .851,  .851,  .851} LT    & \cellcolor[rgb]{ 1,  1,  1}Low & LT    & \cellcolor[rgb]{ 1,  1,  1}Low & LT    & \cellcolor[rgb]{ 1,  1,  1}Low & LT    & \cellcolor[rgb]{ 1,  1,  1}Low & -     & \cellcolor[rgb]{ 1,  1,  1}- & -     & \cellcolor[rgb]{ 1,  1,  1}- \\
    \rowcolor[rgb]{ .851,  .851,  .851} LT    & \cellcolor[rgb]{ 1,  1,  1}Low & LT    & \cellcolor[rgb]{ 1,  1,  1}Low & LT    & \cellcolor[rgb]{ 1,  1,  1}Low & LT    & \cellcolor[rgb]{ 1,  1,  1}Low & -     & \cellcolor[rgb]{ 1,  1,  1}- & -     & \cellcolor[rgb]{ 1,  1,  1}- \\
    \rowcolor[rgb]{ .851,  .851,  .851} LT    & \cellcolor[rgb]{ 1,  1,  1}Low & LT    & \cellcolor[rgb]{ 1,  1,  1}Low & LT    & \cellcolor[rgb]{ 1,  1,  1}Low & LT    & \cellcolor[rgb]{ 1,  1,  1}Low & -     & \cellcolor[rgb]{ 1,  1,  1}- & -     & \cellcolor[rgb]{ 1,  1,  1}- \\
    \bottomrule
    \multicolumn{12}{p{12cm}}{Legend: Assertion Roulette (AR), Eager Test (ET), Empty Test (EpT), Lazy test (LT), Redundant Assertion (RA), Constructor Initialization (CI), Unknown Test (UT), Sensitive Equality (SE). } \\
    \bottomrule
    \end{tabular}%
    \label{tab:severity}%
\end{table}%

We asked the developers about the test smells introduced in their respective projects. We wanted to know whether that test smells affect the maintainability of the test code. We next present the main findings.

Two projects (\texttt{Fatiador} and \texttt{csv2bib}) reported that no test smells found in the projects would affect the maintainability or cause any impact on the test code. Table \ref{tab:MaintenanceImpact} refers to the other four projects, and shows the that test smells \textit{AR} is the test smell that has the most negative impact, followed by the test smell \textit{ET} and \textit{UT}. %

For the \texttt{Curso Massa} project, the developer reported that the \textit{AR} test smells would affect maintainability over time. Conversely, for \textit{ET} and \textit{LT} test smells, the developer reported that they do not affect the maintainability of the test code.
Thus, for the \texttt{Curso Massa}, \texttt{l2jserver}, \texttt{Dependency Injection} projects, the developers reported that the \textit{AR} test smells have a negative impact on the test code. For the \texttt{l2jserver} and \texttt{Dependency Injection} projects, the developers reported that the \textit{ET} test smells have a negative impact on the test code. Regarding the \texttt{Our Digital Bank} project, the developer considered that only the \textit{UT} negatively impacts the test code.

Although they have informed that few test smells might negatively impact the code, 5 out of 6 developers reported that performing test smells maintenance in the test code would improve the comprehension and evolution of the system.

\begin{table}[htbp]
  \scriptsize
  \centering
  \def\arraystretch{1.2}
  \caption{Maintainability and impact of test smells (TS) per project}
    \begin{tabular}{ccc ccc ccc ccc}
     \midrule
     \multicolumn{3}{c}{\textbf{CursoMassa}} & \multicolumn{3}{c}{\textbf{l2jserver}} & \multicolumn{3}{c}{\textbf{Dep. Injection}} & \multicolumn{3}{c}{\textbf{Our Digital Bank}} \\
    \midrule
    \textbf{TS} & \textbf{M} & \textbf{I} & 
    \textbf{TS} & \textbf{M} & \textbf{I} &
    \textbf{TS} & \textbf{M} & \textbf{I} &
    \textbf{TS} & \textbf{M} & \textbf{I} 
    \\
    \midrule
    \rowcolor[rgb]{ .851,  .851,  .851} AR & \multicolumn{1}{c}{\cellcolor[rgb]{ 1,  1,  1}Yes} & \multicolumn{1}{c}{\cellcolor[rgb]{ 1,  1,  1}Negative} & \multicolumn{1}{c}{AR} & \multicolumn{1}{c}{\cellcolor[rgb]{ 1,  1,  1}Yes} & \multicolumn{1}{c}{\cellcolor[rgb]{ 1,  1,  1}Negative} & \multicolumn{1}{c}{AR} & \multicolumn{1}{c}{\cellcolor[rgb]{ 1,  1,  1}Yes} & \multicolumn{1}{c}{\cellcolor[rgb]{ 1,  1,  1}Negative} & \multicolumn{1}{c}{AR} & \multicolumn{1}{c}{\cellcolor[rgb]{ 1,  1,  1}No} & \multicolumn{1}{p{4.145em}}{\cellcolor[rgb]{ 1,  1,  1}\centering -} \\
    \midrule
    \rowcolor[rgb]{ .851,  .851,  .851} \multicolumn{1}{c}{AR} & \multicolumn{1}{c}{\cellcolor[rgb]{ 1,  1,  1}Yes} & \multicolumn{1}{c}{\cellcolor[rgb]{ 1,  1,  1}Negative} & \multicolumn{1}{c}{AR} & \multicolumn{1}{c}{\cellcolor[rgb]{ 1,  1,  1}Yes} & \multicolumn{1}{c}{\cellcolor[rgb]{ 1,  1,  1}Negative} & \multicolumn{1}{c}{AR} & \multicolumn{1}{c}{\cellcolor[rgb]{ 1,  1,  1}Yes} & \multicolumn{1}{c}{\cellcolor[rgb]{ 1,  1,  1}Negative} & \multicolumn{1}{c}{AR} & \multicolumn{1}{c}{\cellcolor[rgb]{ 1,  1,  1}No} & \multicolumn{1}{p{4.145em}}{\cellcolor[rgb]{ 1,  1,  1}\centering -} \\
    \midrule
    \rowcolor[rgb]{ .851,  .851,  .851} \multicolumn{1}{c}{AR} & \multicolumn{1}{c}{\cellcolor[rgb]{ 1,  1,  1}Yes} & \multicolumn{1}{c}{\cellcolor[rgb]{ 1,  1,  1}Negative} & \multicolumn{1}{c}{AR} & \multicolumn{1}{c}{\cellcolor[rgb]{ 1,  1,  1}Yes} & \multicolumn{1}{c}{\cellcolor[rgb]{ 1,  1,  1}Negative} & \multicolumn{1}{c}{AR} & \multicolumn{1}{c}{\cellcolor[rgb]{ 1,  1,  1}Yes} & \multicolumn{1}{c}{\cellcolor[rgb]{ 1,  1,  1}Negative} & \multicolumn{1}{c}{EpT} & \multicolumn{1}{c}{\cellcolor[rgb]{ 1,  1,  1}No} & \multicolumn{1}{p{4.145em}}{\cellcolor[rgb]{ 1,  1,  1}\centering -} \\
    \midrule
    \rowcolor[rgb]{ .851,  .851,  .851} \multicolumn{1}{c}{ET} & \multicolumn{1}{c}{\cellcolor[rgb]{ 1,  1,  1}No} & \multicolumn{1}{c}{\cellcolor[rgb]{ 1,  1,  1}\centering -} & \multicolumn{1}{c}{ET} & \multicolumn{1}{c}{\cellcolor[rgb]{ 1,  1,  1}No} & \multicolumn{1}{c}{\cellcolor[rgb]{ 1,  1,  1}\centering -} & \multicolumn{1}{c}{ET} & \multicolumn{1}{c}{\cellcolor[rgb]{ 1,  1,  1}No} & \multicolumn{1}{c}{\cellcolor[rgb]{ 1,  1,  1}\centering -} & \multicolumn{1}{c}{EpT} & \multicolumn{1}{c}{\cellcolor[rgb]{ 1,  1,  1}No} & \multicolumn{1}{p{4.145em}}{\cellcolor[rgb]{ 1,  1,  1}\centering -} \\
    \midrule
    \rowcolor[rgb]{ .851,  .851,  .851} \multicolumn{1}{c}{ET} & \multicolumn{1}{c}{\cellcolor[rgb]{ 1,  1,  1}No} & \multicolumn{1}{c}{\cellcolor[rgb]{ 1,  1,  1}\centering -} & \multicolumn{1}{c}{ET} & \multicolumn{1}{c}{\cellcolor[rgb]{ 1,  1,  1}Yes} & \multicolumn{1}{c}{\cellcolor[rgb]{ 1,  1,  1}Negative} & \multicolumn{1}{c}{ET} & \multicolumn{1}{c}{\cellcolor[rgb]{ 1,  1,  1}Yes} & \multicolumn{1}{c}{\cellcolor[rgb]{ 1,  1,  1}Negative} & \multicolumn{1}{c}{UT} & \multicolumn{1}{c}{\cellcolor[rgb]{ 1,  1,  1}Yes} & \multicolumn{1}{p{4.145em}}{\cellcolor[rgb]{ 1,  1,  1}Negative} \\
    \midrule
    \rowcolor[rgb]{ .851,  .851,  .851} \multicolumn{1}{c}{ET} & \multicolumn{1}{c}{\cellcolor[rgb]{ 1,  1,  1}No} & \multicolumn{1}{c}{\cellcolor[rgb]{ 1,  1,  1}\centering -} & \multicolumn{1}{c}{ET} & \multicolumn{1}{c}{\cellcolor[rgb]{ 1,  1,  1}No} & \multicolumn{1}{c}{\cellcolor[rgb]{ 1,  1,  1}\centering -} & \multicolumn{1}{c}{ET} & \multicolumn{1}{c}{\cellcolor[rgb]{ 1,  1,  1}No} & \multicolumn{1}{c}{\cellcolor[rgb]{ 1,  1,  1}\centering -} & \multicolumn{1}{c}{UT} & \multicolumn{1}{c}{\cellcolor[rgb]{ 1,  1,  1}Yes} & \multicolumn{1}{p{4.145em}}{\cellcolor[rgb]{ 1,  1,  1}Negative} \\
    \midrule
    \rowcolor[rgb]{ .851,  .851,  .851} \multicolumn{1}{c}{LT} & \multicolumn{1}{c}{\cellcolor[rgb]{ 1,  1,  1}No} & \multicolumn{1}{c}{\cellcolor[rgb]{ 1,  1,  1}\centering -} & \multicolumn{1}{c}{LT} & \multicolumn{1}{c}{\cellcolor[rgb]{ 1,  1,  1}No} & \multicolumn{1}{c}{\cellcolor[rgb]{ 1,  1,  1}\centering -} & \multicolumn{1}{c}{LT} & \multicolumn{1}{c}{\cellcolor[rgb]{ 1,  1,  1}No} & \multicolumn{1}{c}{\cellcolor[rgb]{ 1,  1,  1}\centering -} & \multicolumn{1}{c}{\centering -} & \multicolumn{1}{c}{\cellcolor[rgb]{ 1,  1,  1}\centering -} & \multicolumn{1}{p{4.145em}}{\cellcolor[rgb]{ 1,  1,  1}\centering -} \\
    \midrule
    \rowcolor[rgb]{ .851,  .851,  .851} \multicolumn{1}{c}{LT} & \multicolumn{1}{c}{\cellcolor[rgb]{ 1,  1,  1}No} & \multicolumn{1}{c}{\cellcolor[rgb]{ 1,  1,  1}\centering -} & \multicolumn{1}{c}{LT} & \multicolumn{1}{c}{\cellcolor[rgb]{ 1,  1,  1}No} & \multicolumn{1}{c}{\cellcolor[rgb]{ 1,  1,  1}\centering -} & \multicolumn{1}{c}{LT} & \multicolumn{1}{c}{\cellcolor[rgb]{ 1,  1,  1}No} & \multicolumn{1}{c}{\cellcolor[rgb]{ 1,  1,  1}\centering -} & \multicolumn{1}{c}{\centering -} & \multicolumn{1}{c}{\cellcolor[rgb]{ 1,  1,  1}\centering -} & \multicolumn{1}{p{4.145em}}{\cellcolor[rgb]{ 1,  1,  1}\centering -} \\
    \midrule
    \rowcolor[rgb]{ .851,  .851,  .851} \multicolumn{1}{c}{LT} & \multicolumn{1}{c}{\cellcolor[rgb]{ 1,  1,  1}No} & \multicolumn{1}{c}{\cellcolor[rgb]{ 1,  1,  1}\centering -} & \multicolumn{1}{c}{LT} & \multicolumn{1}{c}{\cellcolor[rgb]{ 1,  1,  1}No} & \multicolumn{1}{c}{\cellcolor[rgb]{ 1,  1,  1}\centering -} & \multicolumn{1}{c}{LT} & \multicolumn{1}{c}{\cellcolor[rgb]{ 1,  1,  1}No} & \multicolumn{1}{c}{\cellcolor[rgb]{ 1,  1,  1}\centering -} & \multicolumn{1}{c}{\centering -} & \multicolumn{1}{c}{\cellcolor[rgb]{ 1,  1,  1}\centering -} & \multicolumn{1}{p{4.145em}}{\cellcolor[rgb]{ 1,  1,  1}\centering -} \\
    \midrule
    \multicolumn{12}{l}{Legend: Maintainability (M), Impact (I).} \\
    \bottomrule
    \end{tabular}%
  \label{tab:MaintenanceImpact}%
\end{table}%

\subsection{What is the behavior of test code after refactoring test smells? (RQ2)}

Before the interviews, we ran the JNose Test for each project. The tool generated a \texttt{csv} file containing the test smells identified in the projects. During the interviews, we asked the developers to refactor some test smells and next commit the changed test files to their GitHub repositories. 
~Then, we introduced each test smell and asked if they could refactor them. 
After the modifications, we run the JNose Test for each project again, obtaining a new \texttt{csv} file with the latest results. We next present the results for each project.

\textbf{Dependency Injection.} We selected 9 test smells: 3 \textit{AR}, 3 \textit{LT}, and 3 \textit{ET} to present to the developer. From the three \textit{AR} test smells, he decided to refactor two out of them into two distinct test classes. He successfully removed the test smells from the test code. From the three \textit{LT} test smells, he refactored one smell, but he did not remove it. Particularly, for the \textit{LT} test smell, the developer should place the creation of an object in the \textit{setUp()} method, but he linked the \textit{LT} test smell to the use of external components, and to refactor, he made use of mocks. Hence, he performed incorrect refactoring. We also observed that the LOC and the number of test methods did not change after refactoring this project. The other test classes, methods, and smells remained without changes. 

\textbf{Curso Massa.} We selected 9 test smells: 3 \textit{AR}, 3 \textit{ET} and 3 \textit{LT}. The developer decided to refactor the \textit{AR} ones. From the \textit{AR} test smells refactored by the developer, only one was removed. The remaining \textit{AR} were refactored but not removed. To refactor \textit{AR} smells, it is necessary to include an explanation (message) in the first parameter of the \textit{assertion} (e.g., \texttt{assertTrue([message,] boolean condition})). The developer placed the message as the last parameter instead.\looseness=-1

\textbf{Csv2bib.} We selected 6 test smells: 2 \textit{AR}, 2 \textit{RA}, and 2 \textit{CI}. The developer decided to refactor 1 \textit{CI} test smell. After refactoring, he removed the test smell. The \textit{CI} test smell occurs when a test method implements a constructor. It is necessary to initialize all the fields within the \textit{setUp()} method. Therefore, for that test smell, a possible refactoring would be to remove the constructor method since it had no instructions. However, the developer attempted to refactor the constructor method by including instructions for setting variables used in other test methods. The JNose Test continued to consider it a \textit{CI} test smell. Since the developer modified the test class, the LOC and the number of test methods changed. For the \texttt{Run.java} test class, its LOC increased from 44 to 57, and the number of test methods risen from 2 to 5.

\textbf{Fatiador.} We selected 9 test smells: 3 \textit{SE}, 3 \textit{AR}, and  3 \textit{LT}. The developer decided to refactor 2 \textit{AR} smells from 2 different test classes: \texttt{DecimalWriter.java} and \texttt{IntegerWriter.java}. After refactoring, the developer successfully removed the two smells. For each test class, he added 1 LOC each (from 47 to 48 LOC and from 40 to 41 LOC), respectively. The other test smells did not present any change.\looseness=-1

\textbf{l2jserver2.} We selected 9 test smells: 3 \textit{AR}, 3 \textit{ET}, and 3 \textit{LT}. The developer refactored 2 \textit{AR} and 1 \textit{ET}. In addition, he also refactored other \textit{AR} test smells not previously selected. For example, the \texttt{BitSetIDAllocator.java} test class contained 7 \textit{AR} before refactoring. Although we asked him to refactor just 1 of them, he refactored 3 occurrences of this test smell. In total, he refactored 4 \textit{AR} test smells for 2 different classes (\texttt{BitSetIDAllocator.java} and \texttt{CharacterIDProvider.java}), in which all of them were successfully removed after refactoring. 

For the  \textit{ET} test smell, the developer chose to refactor the smell inside the \texttt{CharacterIDProvider.java} test class. Therefore, it was necessary to remove the multiple calls to the multiple production methods. However, the developer did not perform the necessary removals.  Moreover, during the incorrect refactoring of the \textit{ET} test smell, he introduced a new type of smell in the project, the \textit{LT}, which occurs when multiple test methods call the same production method. Thus, 2 test classes (\texttt{CharacterIDProvider.java} and \texttt{WorldServiceImpl.java}) have changed. The first increased in 1 LOC, while the latter decreased by 1. For the number of test methods and other test smells, there was no change in behavior.

\textbf{Our Digital Bank.} We selected 6 test smells: 2 \textit{AR}, 2 \textit{Ept}, and 2 \textit{UT}. The developer decided to refactor 1 \textit{AR}, 2 \textit{Ept}, and 2 \textit{UT} test smells. This project contained 4 test classes. During the refactoring, the developer removed 2 of them. In addition, the \texttt{Validator.java} test class contained 35 LOC, and 90 LOC after refactoring. Also, the amount of test methods increased from 4 to 8. All the 3 types of test smells, \textit{AR}, \textit{UT}, and \textit{Ept} presented different behaviors after refactoring: the amount of \textit{AR} test smells increased from 2 to 9; the 2 \textit{UT} refactored were successfully removed; and 2 \textit{Ept} smells were refactored and removed. The project had  2 \textit{Ept}, 1 was removed after refactoring and the other after removing one of the test classes. Additionally, after refactoring, the developer added other 9 test smells, 7 \textit{IT} and 2 \textit{LT}. From the analyzed projects, \texttt{Our Digital Bank} was the one that had the most changes in the test code, which had decreased the number of test classes, increased the number of test smells, and the number of LOC and test methods. After refactoring, two test smells that did not exist before in the project were added to the test code. 


\section{Discussion} \label{discussion}

In this study, we considered three degrees of severity (low, middle, and high) for the test smells. We used this approach to evaluate eight test smells (\textit{AR}, \textit{Ept}, \textit{UT}, \textit{ET}, \textit{LT}, \textit{CI}, \textit{SE}, and \textit{RA}). Five test smells were considered as \textit{low} severity for all projects: \textit{LT}, \textit{SE}, \textit{EpT}, \textit{RA}, and \textit{CI}. From the set of test smells the interviewees analyzed (48 test smells identified in their projects), 42 were considered as \textit{low} severity ($\approx$87,5\%). Only 6 out 48 test smells varied in degrees of severity ($\approx$12,5\%), 4 \textit{AR}, 1 \textit{ET}, and 1 \textit{UT}, spread over 4 projects. 

The \textit{AR} was the one that varied the most. It was perceived by the developers in the three severity degrees, while the \textit{ET} and \textit{UT} presented two different degrees. According to the developers, the \textit{AR} test smell might harm test code maintainability. For example, one interviewee 
claimed that \textit{``if someone else is going to test the system, they will not know why the test failed''}. Another interviewee 
also commented about the \textit{UT} test smells, as follows: 
\textit{``UT harms my code because it shows that it has unnecessary code, and triggers a function in my system that is not validated as it should''}.

Although most test smells received a low severity degree, the developers reported that, in general, test smells might be indicative of problems and harmful to the test code. On the other hand, they claimed that their systems were relatively small, containing simple test cases, and therefore, the test smells may not have as much impact. For larger projects, on the other hand, they may affect more. Therefore, there is a need to conduct further studies to know if those and other types of test smells have a degree of severity that would be different.

According to the developers, fixing test smells may positively impact the test code by improving the quality of the test code, making it more comprehensible, and easing the system's evolution. Two out of six developers also believe that refactoring test smells can result in detecting more bugs. Although four developers claim that refactored test smells do not contribute to the detection of bugs. \looseness=-1

Moreover, for one developer, the \textit{LT} should not be considered a test smell at all. They argue that it might be necessary to have more than 1 test method testing the same production method. In this case, each test method has a different purpose concerning the production class method. For example, for a given test method, the focus may be to exclude something, while in another, it may be to change something else; and the developer cannot see how to perform the test methods in a single method.

\section{Threats to Validity}
\label{threats}

\textbf{Internal validity.} We selected projects from 84 to 2,362 lines of code. This subset of projects may not represent industrial software systems, and therefore replications in this context are desirable. Although Deursen et al. \cite{van2001refactoring} address 21 test smells, this study considered only 8. However, we selected the most frequent test smells in the projects, gathered from the JNose Test results.

\textbf{External validity.} 
The number of participants we found willing to participate in the interview represents a threat to the results. Nonetheless, we sent emails to 90 GitHub developers, and six replied. This sample is not representative, and we expect to replicate this study with a more significant number of participants. Although the results cannot be generalized, they represent a partial view of developers' practices. We believe that this preliminary study is significant and indicative of a trend in the area.  In addition, we intend to investigate this topic further with a more considerable number of test smells and projects. We also present the step-by-step methodology of this work that may allow further replications of this study. Although most of the smells investigated in this study show low severity, we cannot generalize to other projects and other test smells.

\textbf{Construct validity.} 
The interpretation might influence the research results. Regarding refactoring, we did not provide information on how one developer should refactor test smells. We only provided them with definitions and how to detect each of them. The number of refactored test smells varied in each project because we asked the developers if they could refactor and then which one(s) they would like to refactor. The number of smells refactored depended on the developer's choice. As each project presented different test smells, it was impossible to compare the same test smell in all projects.  Since the projects have not been in development for more than two years, this can threaten validity. However, after refactoring, all projects were committed to Github without any problems.\looseness=-1

\section{Related Work}
\label{relatedWork}

Silva-Junior et al. \cite{silva2020survey} presented a study to understand whether test professionals insert unintentional test smells. They surveyed experts to analyze the frequency of use of a set of test smells during the creation and execution of the test cases. The survey was conducted with 60 professionals from different companies and approached 14 test smells widely studied in the literature. According to the authors, experienced professionals introduce test smells during their daily programming activities; they also can provide insights for a better comprehension of how and what practices can lead to the insertion of test smells in the test code. Our study extends such study by providing further information about the degree of severity of test smells and the consequence of refactorings from a developers' perspective.

Spadini et al. \cite{spadini2020investigating} present an investigation into the severity classification concerning four types of test smells and their impact on maintaining the set of tests implemented by the developers. The authors analyzed about 1,500 open source projects to obtain thresholds for the test smells' severity. The authors also integrated a tool for detecting test smells (\textit{tsDetect}) in a prototype extension (back-end) of the \textit{BetterCodeHub} (BCH), a web-based code quality analysis tool. In this study, 31 developers of the BCH project interacted with the prototype. They had to comment on the instances of test smells in the code base of 47 projects. As a result, the developers detected 301 test smells. In the developers' perception, \textit{EpT}, \textit{Sleepy Test}, and \textit{Mystery Guest} present the highest priority as candidates for refactorings, while \textit{Empty Test,} \textit{Ignored Test}, and \textit{Conditional Test Logic} were considered the smells with the most significant impact on the code maintenance. Unlike such study \cite{spadini2020investigating}, in this work, we investigated the severity of 8 test smells through three degrees of severity and refactored the test smells from the developers' perspective. To support the detection of the test smells, we used the JNose Test \cite{Tassio2020Tools}.

Bleser et al. \cite{de2019assessing} conducted two empirical studies. The first observed the propagation of test smells on 164 open source SCALA projects available on GitHub. The authors implemented a tool called \textit{SOCRATES} for automated test smells detection. In the second study, they analyzed the perception and capability of 14 developers in SCALA projects to identify test smells. As a result, test smells had a low propagation among test classes. The most frequently test smells were \textit{Lazy Test}, \textit{Eager Test}, and \textit{Assertion Roulette}. 
%
~Our work differs from such a study as we analyzed developers' perceptions regarding the degree of severity and the consequences of refactoring smells tests. 

\section{Concluding Remarks}
\label{conclusion}

This study aimed to analyze how developers perceive the severity of test smells on test code quality. To accomplish this goal, we selected a set of open-source software projects from Github and interviewed their developers. We also asked them to refactor their test code to remove test smells. Hence, we could observe, from their standpoint, how the developers could consider the effects of such removal.  

The results indicated that test smell severity is highly dependent on the context. For example, for the same project, a given test smell may present different degrees of severity. Regarding test smell refactoring, our initial results pointed out that testers may not know how to refactor a test smell to remove it. In addition, we also identified that it is likely that the removal process could induce the inclusion of new test smells.

As future work, we plan to replicate this study with larger projects, including a more extensive set of test smells. We also plan to investigate the consequences of refactoring test smells further.

\subsubsection*{Acknowledgments.}
This research was partially funded by INES 2.0; CNPq grants 465614/2014-0 and 408356/2018-9 and FAPESB grants BOL0599/2019 and JCB0060/2016.

\bibliographystyle{splncs04}
\bibliography{ref}

\end{document}